\documentclass[12pt,a4paper,twoside]{article}
\usepackage{epsfig}
\newcommand{\bw}{\mathbf{w}}
\newcommand{\qvev}{\langle q \rangle}

\newcommand{\be}{\begin{equation}}
\newcommand{\ee}{\end{equation}}
\newcommand{\balpha}{\mbox{\boldmath$\alpha$}}

\newcommand{\bb}{\mbox{\boldmath$b$}}
\expandafter\ifx\csname mathbbm\endcsname\relax

\else

\fi
\begin{document}
\begin{titlepage}
\begin{flushleft}  
       \hfill                      {\tt hep-th/9603082}\\
       \hfill                      UUITP-4/96\\
       \hfill                       March 1996\\
\end{flushleft}
\vspace*{3mm}
\begin{center}
{\LARGE Notes on Equivalences and Higgs Branches\\
in N=2 Supersymmetric Yang-Mills Theory \\}
\vspace*{12mm}
{\large Ulf H. Danielsson\footnote{E-mail: ulf@teorfys.uu.se} \\
\vspace*{5mm}
P\"{a}r Stjernberg\footnote{E-mail: paer@teorfys.uu.se}\\
\vspace{5mm}
{\em Institutionen f\"{o}r teoretisk fysik \\
Box 803\\
S-751 08  Uppsala \\
Sweden \/}\\}
\vspace*{15mm}
\end{center}

\begin{abstract}
In this paper we investigate how various equivalences between effective field theories of $N=2$ SUSY Yang-Mills theory with matter can be understood through Higgs breaking, i.e. by giving expectation values to squarks. We give explicit expressions for the flat directions for a wide class of examples.
\end{abstract}

\end{titlepage}

\section{Introduction}

After the initial work of N. Seiberg and E. Witten \cite{SW} many $N=2$ SUSY gauge theories with different kinds of matter have been solved. The classical groups with matter in the fundamental representation are all well understood, [2-13], while examples with other matter representations, and in particular exceptional gauge groups, are more scattered \cite{var2,AAM}.

In this note we will consider the Coulomb and Higgs branches of several different $N=2$ Yang-Mills theories with various matter content.  We will build on the analysis of
\cite{var2} where several different combinations of gauge groups and matter hypermultiplets were discussed. In that paper it was shown how seemingly very different theories had the same effective behavior for the massless $U(1)$ fields.
The way to find equivalent theories is to find groups where the weights of some matter representations partly overlap the root system. This leads to cancellations in the prepotential. The remaining roots and weights may then form the adjoint representation and some new matter representation for another group.
What happens is that the 1-loop contribution from the matter hypermultiplets cancel the contributions from some of the W-bosons and we effectively end up with another gauge group. Following \cite{SW} one can then argue that also the nonperturbative structure of the two theories is the same.

In \cite{AS} another way to understand some of these equivalences were discussed. If the squarks in the hypermultiplets are given expectation values in some flat direction this will in general break the gauge symmetry. In addition the hypermultiplet (or parts of it) is lifted and becomes massive. The subject of the present paper is to understand the equivalences discovered in \cite{var2} in this physical setting.

The reasoning of \cite{AS} is valid for large squark expectation values where the theory is weakly coupled. However, due to a nonrenormalization theorem discussed in \cite{AS,APS2}  nothing happens to the effective theory even as we tune the expectation values all the way down to zero. At zero we reach the situation discussed in \cite{var2}. This is where the Higgs branch touches the Coulomb branch. The cancellation mechanism used in \cite{var2} gives an illustration of the nonrenormalization theorem at work.

In section two we will briefly review the equations determining the flat directions. We will also describe how group theory tools can be efficiently used to obtain flat directions also for exceptional groups.
In \cite{var2} three classes of equivalences were given based on group theory reasoning. In section three we will discuss these three classes in turn and show that they can all be understood through Higgs breaking. The analysis performed in \cite{var2} can in fact be seen as an efficient way of determining what kind of flat directions there are. Finally we end with some comments on a case that do not fit into the above picture.

\section{How to find flat directions}

\subsection{Equations for flat directions}

We will consider an $N=2$ SUSY gauge theory with gauge group $G$ and with some number of matter hypermultiplets in the representations $R_i$. In $N=1$ language the $N=2$ vector multiplet includes a gauge field multiplet  $W=(A, \lambda )$ and a chiral multiplet $\Phi =(\phi , \psi)$. $A$ is a vector, $\lambda$ and $\psi$ are Weyl fermions, while $\phi$ is a complex scalar, all in the adjoint representation. The matter multiplet consists of two chiral multiplets $Q= (q,\psi )$ and $\tilde{Q} =(\tilde{q}, \tilde{\psi})$, both with a complex scalar and a Weyl fermion. $q$ is the scalar squark while $\tilde{q}$ is the scalar antisquark.

In addition to these physical fields we also have, off shell, the auxiliary fields.
The scalar potential of $N=2$ SUSY Yang-Mills is obtained by integrating out these auxiliary fields using their equations of motion. In the $N=2$ vectormultiplet we find the fields $D$ and $f$, where  $D$ is real and $f$ is complex. In $N=1$ language the $D$ is sitting in the $N=1$ vectormultiplet while $f$ comes from the adjoint matter. These three (real) auxiliary fields build up an $SU(2)_{R}$ triplet.

Following \cite{PW} we find in this way a contribution to the potential of the form:
\be
(q_i^{\dagger} T^{a} q^i - \tilde{q_i} T^{a} \tilde{q}^{i \dagger})^2 + 4 (\tilde{q_i} T^{a} q^i )^2.
\label{pot}
\ee
$T^a$ are the generators of the gauge group where $a=1,...,\mbox{dim} G$.
The first term is due to the $D$ field, the second is due to the $f$. That is, it comes from the superpotential
\begin{equation}
W=\sqrt{2}\tilde{Q_i} \Phi Q^i+\sum_{i}^{}{m_j^i \tilde{Q}_i Q^j}.
\end{equation}
The relative coefficient in (\ref{pot}) is fixed by the $N=2$ $SU(2)_R$.

The flat directions are obtained by requiring each of the individual contributions to the potential (all positive semidefinite) to vanish. From (\ref{pot}) we find
\begin{equation}\label{D2}
q_i^\dagger T^a q^i - \tilde{q}_i  T^a \tilde{q}^{i \dagger}   = 0
\end{equation}
and
\be
\tilde{q_i} T^{a} q^i =0.  \label{f}
\ee
Further positive semidefinite contributions to the potential are obtained by integrating out the complex auxiliary fields $F$ and $\tilde{F}$ of the $N=2$ matter.
The $F$ term equations require
\begin{equation}\label{F2}
{F_i^a}^\dagger = - {\partial {W}\over \partial  q_a^i } =- \sqrt{2} \phi_b^a \tilde{q}_i^b  - m_i^j \tilde{q}_j^a =0
\end{equation}
\begin{equation}\label{F1}
\tilde {F}_i^{a \dagger}=-{\partial {W}\over \partial \tilde{q}_i^a } =-\sqrt{2} \phi_a^b q_b^i - m_j^i q_a^j =0.
\end{equation}
Finally we also have
\begin{equation}\label{D1}
\mbox{Tr} [\phi^\dagger,\phi]^2=0.
\end{equation}
These, then, are the equations that we need to solve.

\subsection{Solving the equations}

To start with, the generic solution of (\ref{D1}) can be written $\phi =\sum _i b_i H^i$, where $H_i$ are the generators of the Cartan subalgebra. It has become standard to refer to the subspace  of the moduli space where $q=\tilde{q} = 0$ as the Coulomb branch. The subspace $\phi =0$ is called the pure Higgs branch. Finally, on a mixed branch both $q$ or $\tilde{q}$  and $\phi$ have non-zero vacuum expectation values.

Flat directions and symmetry breaking are conveniently studied in terms of directions in weight space. In the case of exceptional groups this technique is especially useful and leads to substantial simplifications.
Suppose that the matter multiplets are in a representation $R$ of the gauge group. The squark fields can then be written as

\begin{equation}
q^i=\sum{q^i_a | \mathbf{w}^a \rangle},
\end{equation}
where $i$ is a flavor index and $a=1,...$dim$R$. $| \mathbf{w}^a \rangle$ is the direction in the Hilbert space corresponding to the weight $\mathbf{w}^a$. The effect of giving a VEV to the squark fields in a flat direction  is that some gauge bosons acquire a mass from the minimal coupling term

\begin{equation}
g^2 A_\mu^a A_\nu^b q^\dagger T_a^\dagger T_b q  .
\end{equation}

If the generators are written in the Cartan-Weyl basis we can read off in the root and weight diagrams which gauge bosons become massive. In fact, since supersymmetry is unbroken the whole $N=2$ vector multiplet is lifted. Furthermore, when the scalar fields are translated the quartic terms in (\ref{pot}) give rise to mass terms for the squarks and antisquarks. By $N=2$ supersymmetry the whole hypermultiplet becomes massive.

The group theoretical resoning in \cite{var2} gave a list of examples with either one or two flavors. The strategy to find flat directions in the different cases is the following:
\begin{enumerate}
\item If we have a pair of hypermultiplets in a complex representation, then the equations (\ref{D2}) are satisfied by taking $\langle q^1 \rangle = v | \bw \rangle$ and  $\langle \tilde{q}_1 \rangle = 0$ for the first flavor and $\langle \tilde{q}_2 \rangle = v | -\bw \rangle$ and    $\langle q^2 \rangle = 0$ for the second flavor. $\bw$ is a weight in $R$ and $-\bw$ a weight in $\bar{R}$. A squark and an antisquark can not acquire expectation values in the same flavor in order for (\ref{f}) to be satisfied. Satisfying (\ref{F2},\ref{F1}) requires a fine tuning of the Higgs vacuum expectation values that reduces the rank.

\item If the representation is real $\bw$ and $-\bw$ are in the same representation. This makes it possible to take either $\langle \tilde{q}_2 \rangle = v | - \bw \rangle$ or $\langle q_2 \rangle = v | -\bw \rangle$ while keeping $\langle q_1 \rangle = v | \bw \rangle$. Again the $F$-term equations require a fine tuning that lower the rank.

\item If there is only one flavor, we try to find a linear combination  $\qvev = \sum_{a=1}^{n}{|\bw^a \rangle}$ such that the sum over each individual component of the weight vectors vanishes, i.e. $\sum_{a=1}^{n}{{(\bw^a)}^j} = 0$. We also require $\langle \tilde{q} \rangle =0$. This guarantees that $\langle q^\dagger \rangle H^i \langle q \rangle = 0$. If there are no root vectors that connect two of the weights in the linear combination then $\langle q^\dagger \rangle E_{\alpha} \langle q \rangle = 0$  also holds. As above the rank is in general lowered by fine tuning. A special case, however, is when there is a weight $\bw_0$ with all components equal to zero, in which case we can take $\qvev = |\bw_0 \rangle, \ \langle \tilde{q} \rangle = 0.$ In this case the rank is not lowered.
\end{enumerate}

These three ways of solving the equations for the flat directions are sufficient to cover the equivalences discussed in \cite{var2}. In the next section we will provide several examples of this general structure.

\section{Examples}

Subgroups can be divided into regular subgroups, whose root system is a subset of the root system of the full group and special subgroups, see e.g. \cite{S,F}. Regular subgroups fall into two classes: those that have the same rank as the full group and those that have lower rank. Special subgroups always have lower rank. We now give explicit examples from these three classes, using the techniques of the previous section.

\subsection{Regular subgroups with equal rank}

In \cite{var2} the following list of rank preserving equivalences involving regular subgroups was given:
\begin{equation}
	\begin{array}{ccccc}
		 & SO(2n) & \subset & SO(2n+1) & R_M = {\bf (2n+1)}
		\label{RegularSeries}  \\
		 & SU(3) & \subset & G_2 & R_M = {\bf 7}
		\label{RegularG_2}  \\
		\tilde{R}_M = {\bf 9} & SO(9) & \subset & F_4 & R_M = {\bf 26}
		\label{RegularF_4} 		\label{RegularUE_8}
	\end{array} \quad .
\end{equation}

We begin by considering the exceptional example $G_2$ with one hypermultiplet in the fundamental {\bf  7} breaking down to $SU(3)$. The root diagram and  fundamental weight diagram of $G_2$ are shown on the left of fig. 1.  If we want to preserve the rank the only direction along which we can give the squarks a VEV is $\bw ^1 = (0,0,0)$. Otherwise some Cartan element acting  on $\langle q \rangle$ would give a non-zero result. This would give mass to a gauge boson in the Cartan subalgebra and lower the rank of the group. The residual symmetry group is generated by the roots that annihilate $\bw ^1$, since the corresponding group element leave $\langle q \rangle$ invariant. As is obvious from the figure, it is the long roots that do not touch $\bw ^1$.
The  long roots (plus the two Cartan elements) are then seen to form the adjoint of $SU(3)$.
The six short roots, on the other hand, will correspond to massive vectors.  Turning to the matter multiplet,  we recognize in the weight diagram the fundamentals {\bf 3} + {\bf \={3}} and one singlet of $SU(3)$.
We shift the scalar fields $q \rightarrow  q_1 + v,q_a  \rightarrow  q _a, a=2,...,7$ and write
\begin{equation}
q = (q_1 + v) | \mathbf{w}^1 \rangle + \sum_{a=2}^{7}{q_a | \mathbf{w}^a \rangle}.
\end{equation}
We then see that a squark field gets mass from the $D$-term  if it can be rotated by a ladder operator $E_\alpha$ to lie in the direction of $\qvev$. In this particular case $q_2,...,q_7$ obtain a mass $\sim |a|^2$. Similarly $\tilde{q}^2,...,\tilde{q}^2$ get mass from the $f$-term. Note that the factor 4 in front ensures that the antisquarks get the same mass as the squarks.
The singlet will remain massless and corresponds to the flat direction.

\begin{figure}                                                           
\epsfig{file=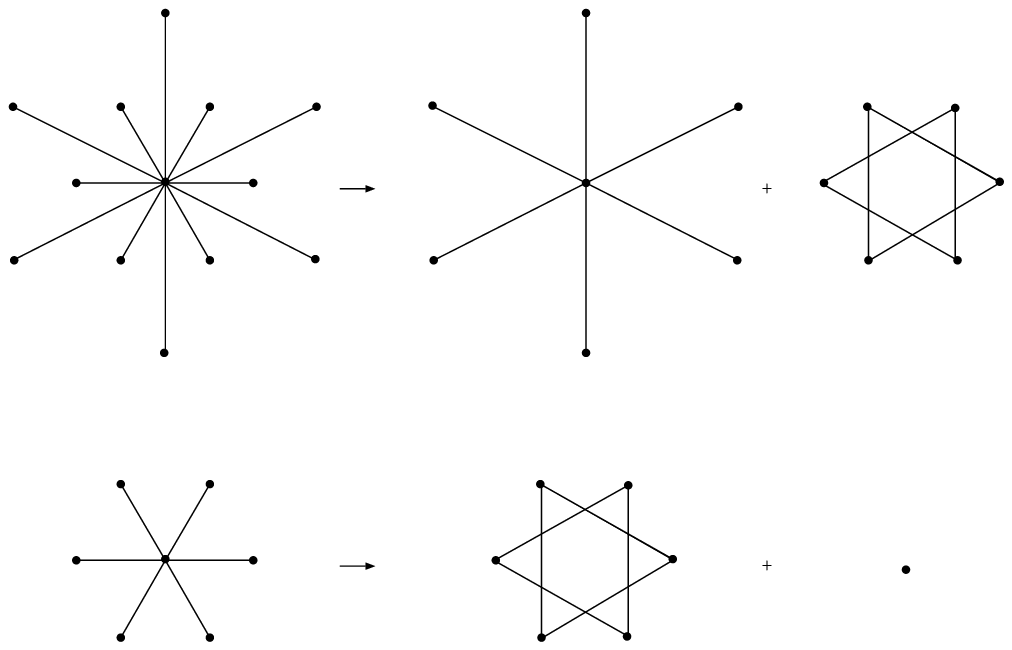,width=15 cm}                                               
\caption{$G_2 \rightarrow SU(3)$ with ${\bf  14} \rightarrow {\bf 8} + {\bf 3} + {\bf \bar{3}}$ and ${\bf 7 } \rightarrow  {\bf 3} + {\bf \bar{3}} + {\bf 1}$ .}
\end{figure}

$SO(2n+1) \rightarrow SO(2n)$ can be shown to work out in exactly the same way, with an expectation value for the $\bf 2n+1$ multiplet in the direction $(0,...,0)$.

The last example is a little bit more complicated.
When the exceptional group $F_4$ is broken by a {\bf 26} down to $SO(9)$, one must note that the  {\bf 26} contains two weights with eigenvalues $(0,0,0)$.  Choose $\qvev$ in the direction of one of them, say $\bw$. One can show that there are 16 roots that do not annihilate $\bw$ and the corresponding gauge bosons therefore obtain masses. Clearly there are also 16 weights in the $\bf 26$ that can be reached  from $\bw$ using these roots and the corresponding squarks also get massive. There are 9 weights which can not be reached in this manner, including the other $(0,0,0)$. These build up the ${\bf 9}$ of $SO(9)$. Finally, the hypermultiplet belonging to $\bw$ becomes a massless singlet. All this fits with the branching rule that states ${\bf 52} \rightarrow {\bf 36} + {\bf 16}$ and ${\bf  26} \rightarrow {\bf 1 } + {\bf 9} + {\bf 16 }$.

In all these cases with rank preserving breaking the bare mass of the quarks must be zero for the $F$-term equations to be obeyed.

\subsection{Regular subgroups with lower rank}

The list  of examples from \cite{var2} is
\begin{equation}
\begin{array}{ccccc}
		& SU(n-1) & \subset & SU(n) & R_M ={\bf n}+{\bf \bar{n}} \\
    	& SO(2n-1) & \subset & SO(2n+1) & R_M = 2\cdot {\bf (2n+1)} \\
	    & Sp(2n-2) & \subset & Sp(2n) & R_M =2\cdot {\bf (2n)} \\
	    & SO(2n-2) & \subset & SO(2n) & R_M =2\cdot {\bf (2n)} \\
\tilde{R}_M =2\cdot {\bf 10} & SO(10) & \subset & E_6 &
R_M={\bf 27}+{\bf \bar{27}} \\
	    & E_6 & \subset & E_7 & R_M ={\bf 56}\\
	    & SU(2) & \subset & G_2 &  R_M =2\cdot {\bf 7}\\
  	    & SU(4) & \subset & SO(8) &  R_M =2\cdot {\bf 8}_s \\
\tilde{R}_M={\bf 5}+{\bf \bar{5}}
        & SU(5) & \subset & SO(10) &  R_M={\bf 16}+{\bf \bar{16}} \\
	    & SU(6) & \subset & SO(12) &  R_M={\bf 32}' \\
  	    & SU(2) & \subset & Sp(4) & R_M=2\cdot {\bf 5}  \\
	    & SU(3) & \subset & Sp(6) & R_M={\bf 14}'
  \end{array}
\end{equation}

When the rank of the group is lowered (\ref{F2}) and (\ref{F1}) can only by satisfied if the
VEVs of the Higgs fields are adjusted. If a squark get an expectation
value in some direction $|\mathbf{\Lambda} \rangle$ then (\ref{F1})  gives $\sum _i b_i H^i 
|{\bf \Lambda} \rangle =  \bb \cdot {\bf \Lambda} = 0$ if the bare masses are zero. In the rank lowering breaking in this and the next subsection the bare mass of the quarks can also be nonzero and still allow for solutions of the $F$-term equation, but we will only consider cases with $m_i=0$. This equation specifies a $(r-1)$-dimensional subspace of the Coulomb moduli space, where it is possible to turn on the squark expectation values and switch to a mixed branch of the theory. In general, if  $\qvev$ is a linear combination containing $k$ linearly independent weights ${\bf \Lambda}^1, ...,{\bf \Lambda}^k$, then we get $k$ equations, since  $\sum _i b_i H^i$ acting on each of the state vectors $|{\bf \Lambda}^k \rangle$ has to vanish. In those cases the rank is lowered by $k$.
Let us introduce new coordinates \{$\tilde{b_i}$\} on the reduced moduli space.
The mass squared of a gauge boson is given by $(\bb \cdot \balpha )^{2}$.
The root system of the residual symmetry algebra is obtained by demanding  $\bb \cdot
\balpha = \tilde{\bb} \cdot \tilde{\balpha}$, where $\tilde{\balpha}$ is a root of the smaller
algebra.

Most examples in the list above start with a theory with two hypermultiplets. As discussed earlier a flat direction is given by $\langle q^1 \rangle = v |\bw \rangle$  and $\langle \tilde{q}_2 \rangle = v| - \bw \rangle$. One now has to check if the breaking patterns in the list can be reproduced.
Let us consider one example, the breaking $Sp(6) \rightarrow Sp(4)$, in some detail. All  examples are constructed in such a way that  the Dynkin index are the same for the two theories. In this case the Dynkin index requires two matter multiplets in the fundamental {\bf 6} of $Sp(6)$.  It can also be seen from the fundamental weight diagram that there are no flat directions with only one flavor, because the only way to satisfy (\ref{F2}) is that  $\qvev$ contains terms of the form $ |\bw \rangle + | - \bw \rangle$, but each such pair is interconnected by a root vector as can be checked from the adjoint representation.

Following the prescription given earlier we let $\langle q^1 \rangle = v |1 \ 0 \ 0 \rangle$  and $\langle \tilde{q}_2 \rangle = v| - 1 \ 0 \ 0 \rangle$, where the weights are given in the Dynkin basis. Altogether there are ten roots that annihilate $(1,0,0)$ or $(-1,0,0)$. These are the generators of the the residual symmetry group $Sp(4)$. The remaining roots that become massive arrange into two {\bf 4} and  three singlets.

\begin{figure}  
\begin{center}                                                         
\epsfig{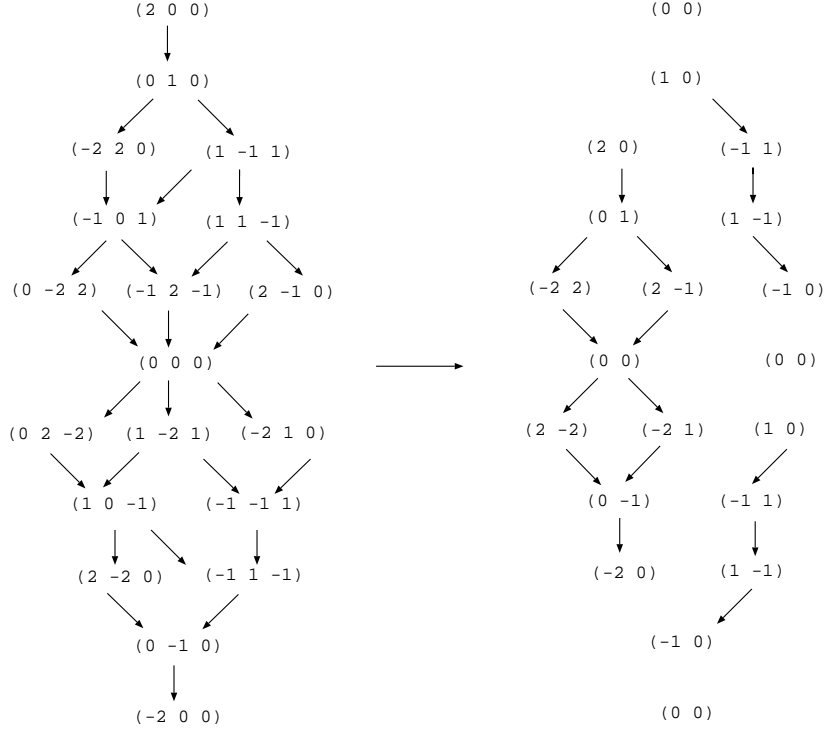}                                               
     
\caption{$Sp(6) \rightarrow Sp(4)$ with ${\bf  21} \rightarrow {\bf 10} + 2 \cdot {\bf  4} + 3 \cdot {\bf  1}$.}
\label{sp1}
\end{center}
\end{figure}

\begin{figure}
 \begin{center}                                                         
\epsfig{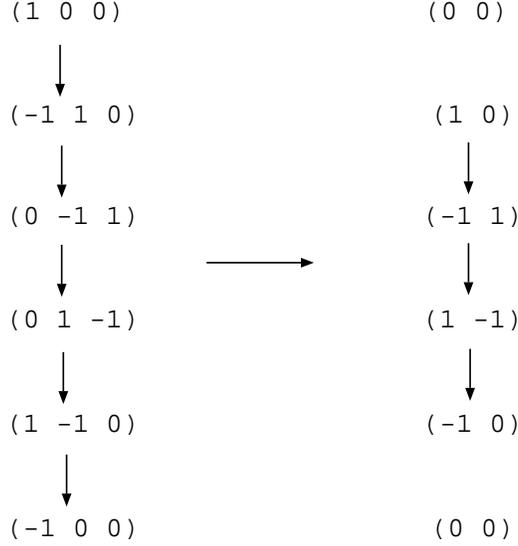}                                               
     
\caption{$Sp(6) \rightarrow Sp(4)$ with ${\bf  6} \rightarrow {\bf 4} + 2 \cdot {\bf  1} $.}

\label{sp2}
\end{center}
\end{figure}

The fine tuning condition is $b_1=0$. We then introduce new coordinates $\tilde{b}_1$ and $\tilde{b}_2$ through $b_1=0,b_2=\tilde{b}_1,b_3=\tilde{b}_2$. The projection induced by requiring $\bb \cdot
\balpha = \tilde{\bb} \cdot \tilde{\balpha}$ is $(a_1,a_2,a_3) \rightarrow (a_2,a_3)$, i.e. the projection just eliminates the first Dynkin label.  The projected root diagram in the first case is shown in fig. \ref{sp1}, where the {\bf 10} and two {\bf 4}'s plus three singlets of $Sp(4)$ can be recognized. The fundamental weight diagram is mapped onto a {\bf 4} and two singlets (fig. \ref{sp2}). The two other cases, $\langle q^1 \rangle = v | -1 \ 1 \ 0 \rangle$ and $\langle q^1 \rangle = v | 0 \ -1 \ 1 \rangle$, give rise to the projections $(a_1,a_2,a_3) \rightarrow (a_1 + a_2,a_3)$ and $(a_1,a_2,a_3) \rightarrow (a_1,a_2 + a_3)$ respectively. In these two cases the projected diagrams are just permutations of the first one.

There are also  three cases with only one hypermultiplet.  All of these matter representations are real (or pseudoreal): ${\bf 56}$ of $E_7$, ${\bf 32'}$ of $SO(12)$and ${\bf 14'}$ of $Sp(6)$. As such they have the property that if  $\bw$ is a weight, then so are $-\bw$.  A flat direction is given by $| \bw \rangle + |- \bw \rangle$, provided there are no root $\balpha$ such that $  -\bw  + -\balpha = \bw$, i.e. $\balpha = 2 \bw$. It can be checked that there are no such roots in any of the three cases by writing down the adjoint representations.

\subsection{Special breaking}

The list from \cite{var2} is
\begin{equation}
\begin{array}{ccccc}
	  & SO(2n-1) & \subset & SO(2n) & R_M={\bf 2n}\\
      & Sp(2n) & \subset & SU(2n) & R_M={\bf n (2n-1)}\\
      & G_2 & \subset & SO(7)  & R_M={\bf 7} {\rm \ or\ } {\bf 8}\\
      & F_4 & \subset & E_6 & {\bf 27}
\end{array}
\end{equation}
 From the first infinite series we consider the example $SO(6)+${\bf 6}$\rightarrow SO(5)$. We can satisfy (\ref{D2}) by taking
\begin{equation}\label{}
\langle q \rangle=v \left( |\bw\rangle + |-\bw\rangle \right)
\end{equation}
for some weight $\bw$.
Since there are no ladder operator $E_\alpha $ that connects a pair of states $|\pm \bw \rangle$, $\langle q_i^\dagger \rangle E_\alpha  \langle q^i \rangle =0$ holds. The $F$-term eq.
\begin{equation}\label{finetunecond}
\sum_{i}{b_i H_i} (|\bw\rangle + |-\bw\rangle) =0,
\end{equation}
gives a fine-tuning condition for the Higgs VEVs. If we choose e.g. $\bw =(-1,0,1)$ then (\ref{finetunecond}) gives $-b_1+b_3=0$. Defining $b_1= \tilde{b_1}$, $b_2= \tilde{b_2}$, $b_3 =  \tilde{b_1}$, the projection $\bb \cdot \balpha = \tilde{\bb} \cdot \tilde{\balpha}$ give rise to the splitting ${\bf  15} \rightarrow {\bf 10} + {\bf 5}$ for the adjoint and ${\bf  6} \rightarrow {\bf 5 } + {\bf 1}$ for the fundamental.
Since the weight diagram for the ${\bf 6}$ of $SO(6)$ are completely symmetric with respect to pairs of weights $(\bw,-\bw)$ (the weights sit in the corners of an octahedron),  the same symmetry breaking pattern is obtained if another pair of weights are chosen for the symmetry breaking direction. Again it is easy to check that the matter multiplets get mass in the expected way from the $D$-term and the $f$-term. The field in the flat direction becomes a massless gauge singlet, while the field in the direction  $|\bw\rangle - |- \! \bw\rangle$ go into the massive ${\bf  5}$. The other cases in the series work out in the same way.

The next infinite series contains matter in nonfundamental representations. As an example, let us consider the case  $SU(6)$ with one  ${\bf 15}$ hypermultiplet, i.e. a complex tensor representation. Contrary to previous cases involving complex representations there is only one matter multiplet. We can choose 
\begin{equation}
\qvev \sim |0 \  - \!1 \ 0 \  1 \ 0  \rangle + |-\!1 \  1 \ 0 \ -1 \ 1  \rangle + |1 \  0 \ 0  \ 0 \ - \!1 \rangle,
\end{equation}
as a flat direction,
where it can be checked that no roots connect the weights. The gauge bosons corresponding to the Cartan elements $H_3$,  $H_2 + H_4$ and $H_1 + H_5$ remain massless, while $H_2 - H_4$ and $H_1 - H_5$ become massive.

We now turn to the two exceptional cases.
Breaking $SO(7)$ to $G_2$ by a ${\bf 8}$ is very similar to $SO(6)+${\bf 6}$\rightarrow SO(5)$. With a hypermultiplet in the {\bf 7}, on the other hand, there does not seem to exist a straightforward implementation of the breaking. The reason for this is that there must always remain a massless singlet of the residual symmetry group, namely the field in the flat direction. The best we can do is to put $\qvev$ in the $(0,0,0)$ direction and break to $SO(6)$. A further, voluntary restriction of the Higgs expectation values then give pure $G_2$. This was the technique used in \cite{var2} to solve $G_2$.

 A more complicated case is the special breaking of $E_6$ by a fundamental ${\bf 27}$ to $F_4$ where the rank is lowered by two. The breaking is achieved by choosing
\begin{equation}
\qvev \sim |0 \  1 \ 0 - \! 1 \ 0 \ 0 \rangle + |1 \  -1 \ 0 \ 1 \ -1 \ 0 \rangle + |-1 \  0 \ 0  \ 0 \ 1 \ 0 \rangle.
\end{equation}
It can be checked that there are no roots that take one of the weights into the other. The gauge bosons corresponding to the Cartan elements $H_3$, $H_6$, $H_2 + H_4$ and $H_1 + H_5$ remain massless, while $H_2 - H_4$ and $H_1 - H_5$ become massive.

\section{Conclusions}

In this paper we have seen how the equivalences discussed in \cite{var2} can be understood through Higgs breaking. We have given explicit solutions for the flat directions in terms of weights in a variety of examples. Clearly the equivalences of \cite{var2} should be understood as due to roots of Higgs branches.

But even if we can physically understand most of the equivalences discussed in \cite{var2}, there are still examples that we have not been able to describe in these terms, e.g.
\be
SU(2n) + 2 \cdot {\bf 2n} \sim SO(2n+1).  \label{konst}
\ee
While we understand cases like e.g. $SU(4) + 2 \cdot {\bf 4} \rightarrow SU(3)$ and $SU(4) +{\bf 6} \rightarrow SO(5)$ the above equivalence does not fit into the picture. In all other cases the hypermultiplet cancels completely the contributions from some gauge bosons. As explained in detail for the case  $n=2$ in \cite{var2}  only a fraction of a given gauge boson contribution is cancelled for (\ref{konst}). This is difficult to understand from the point of view of breaking and lifting. Is the equivalence a coincidence or is there some other physical explanation?

\section*{Acknowledgements}

We wish to thank Bo Sundborg and Gabriele Ferretti for numerous helpful discussions and Philip Argyres for suggesting the problem.

\end{document}